\documentclass[12pt,a4paper]{article}
\usepackage[latin1]{inputenc}
\usepackage[T1]{fontenc}
\usepackage[english]{babel}
\usepackage{lmodern}
\usepackage{hyperref}
\usepackage[dvips]{graphics,color}
\usepackage{relsize}
\usepackage{cancel}
\usepackage{amsmath,amssymb,amsfonts}
\usepackage{bm}
\usepackage{tikz}
\title{Quantum dissipative effects for a real scalar field coupled to a
time-dependent Dirichlet surface in $d+1$ dimensions}
\author{C.~D.~Fosco \\
and \\
B.~C.~ Guntsche\\
{\normalsize\it $^a$Centro At\'omico Bariloche and Instituto Balseiro}\\
{\normalsize\it Comisi\'on Nacional de Energ\'\i a At\'omica}\\
{\normalsize\it R8402AGP Bariloche, Argentina.}}
\begin{document}
\date{}
\maketitle
\begin{abstract}
We study the Dynamical Casimir Effect (DCE) for a real scalar
field $\varphi$ in $d+1$ dimensions, in the presence of a mirror
that imposes Dirichlet boundary conditions and undergoes time-dependent
motion or deformation. Using a perturbative approach, we expand in powers
of the deviation of the mirror's surface $\Sigma$ from a hyperplane, up to
fourth order. General expressions for the probability of pair creation
induced by motion are derived, and we analyze the impact of space-time
dimensionality as well as of the non-linear effects introduced by the
fourth-order terms.	
\end{abstract}
\section{Introduction}\label{sec:intro}
Quantum vacuum fluctuations, in the presence of non trivial boundary
conditions, may induce macroscopic manifestations such as the Casimir and
related effects~\cite{milton1999casimir,bordag2009advances}.  In the
Dynamical Casimir Effect (DCE), particles are created out of the quantum
vacuum, due to the presence of time-dependent boundary conditions. The
latter may adopt different guises, the most prominent among them being a
time-dependence in the geometry of the boundaries, which otherwise keep the
nature of the condition they impose unaltered.  
Within this context, the  creation of particles in an oscillating 
one-dimensional cavity containing a moving mirror is a phenomenon that has been received extensive attention, since the pioneering works by Moore~\cite{Moore}, and
subsequent results by Davies and Fulling~\cite{FulDav}.  

More recent studies have focused on various aspects of the Dynamical Casimir
Effect (DCE), highlighting their relevance to diverse physical phenomena~\cite{reviews}.  In a perhaps less direct way, the DCE may shed light into the information loss problem and even entanglement entropy~\cite{Chen:2017lum}. Also, the study of physical observables in moving mirror models~\cite{Good} sheds light on the physics of evaporating black holes. 
Note that moving mirrors are an example of entanglement harvesting from the vacuum~\cite{Cong}, and that studies have explored their radiation from the point of view of the Equivalence Principle~\cite{Fulling}. Finally, in an apparently rather different context, Dirichlet boundary conditions for a quantum field in higher dimensions appear naturally in the subject of D-branes~\cite{Polchinski:1996na, Johnson:2023onr}.

In this paper, we deal with a massless real scalar field, in $d+1$ dimensions, under the influence of Dirichlet boundary conditions on a space and time 
dependent surface. The goal is to derive expressions for the imaginary part of the real-time effective action resulting from the integration of the scalar field, which is a functional of the surface. We recall the fact that the imaginary part determines the probability of vacuum decay; indeed, denoting by ${\mathcal P}$ that probability, it may be written as ${\mathcal P} = 2 \, {\rm Im} \Gamma$, where $\Gamma$ denotes the effective action. In our case, that probability corresponds to processes whereby the vacuum decays to states containing a certain number of real quanta of the scalar field.

This paper is organized as follows: In Sect.~\ref{sec:thesystem}, we define
the system to be studied in the remainder of the paper and its effective
action. Then, in Sect.~\ref{sec:pert} we focus on the evaluation of the
same object in perturbation theory, discarding whenever possible, pieces
of the effective action which cannot contribute to its imaginary part. 
We present results for both the second and the fourth orders in the expansion.
Finally, in Sect.~\ref{sec:conc} we present our conclusions.

\section{The system and its effective action}\label{sec:thesystem}
The system that we study here, consists of a massless real scalar field $\varphi(x)$
in $d+1$ dimensions, subjected to  Dirichlet boundary conditions on a
space-time surface $\Sigma$, but otherwise being described by a free action
${\mathcal S}_0(\varphi)$:
\begin{equation}\label{eq:defs0}
{\mathcal S}_0(\varphi) \,=\, \frac{1}{2} \int d^{d+1}x \, 
\partial_\mu \varphi(x)  \partial^\mu \varphi(x) \;.
\end{equation}
In order to simplify the perturbative calculations, we adopt
imaginary-time conventions~\footnote{Note, however, that results will be
continued back to real time when evaluating the imaginary part of the
effective action.}, such that the metric becomes Euclidean, i.e.,
\mbox{$[g_{\mu\nu}] = [{\mathbb I}_{\mu\nu}]$}, where ${\mathbb I}$ is the $(d+1)
\times (d+1)$ identity matrix. Space-time indices are, in our conventions,
represented by letters from the middle of the Greek alphabet ($\mu, \nu,
\ldots$) and run over the values $0, 1, \ldots, d$ ($d$ is
the number of spatial dimensions). Space-time coordinates are denoted by
$x^\mu = x_\mu$, $x^0$ being the imaginary time. Einstein convention of summation
over repeated indices in monomial expressions is also assumed. We adopt natural units, such that: $\hbar \equiv 1$ and $c \equiv 1$.  

Note that, had the field been free, i.e., devoid of any non trivial
boundaries, the effective action would have coincided with ${\mathcal
S}_0$. We want, however, to include the effects of a space-time surface,
$\Sigma$, where the field is constrained to vanish.  This will produce,
when taking into account the scalar field fluctuations, a quantum, one-loop
contribution to the effective action, depending just on the surface. We
denote this term as $\Gamma(\Sigma)$, and we may write it in terms of a
functional integral:
\begin{equation}\label{eq:defgamma}
e^{-\Gamma(\Sigma)} \;=\; \frac{1}{\mathcal N} \; 
\int {\mathcal D}\varphi \, \delta_\Sigma(\varphi) \, 
e^{- {\mathcal S}_0(\varphi)} \;\;,\;\;\;
{\mathcal N} \,=\, \int {\mathcal D}\varphi \, e^{- {\mathcal S}_0(\varphi)}
\;.	
\end{equation}
Here, $\delta_\Sigma(\varphi)$ represents a Dirac $\delta$ functional, imposing 
Dirichlet conditions for $\varphi$ on the surface.  An important remark is in order here: we shall be only interested in the imaginary part of $\Gamma(\Sigma)$, after rotating this object back to real time. In particular, this implies that we may safely discard all the local, divergent counterterms that will arise. 

In what follows, we make an assumption about the surface, namely, that it
may be described, at least with a proper choice of coordinate system, by
a single Monge patch. More concretely, letting 
$x_\shortparallel \equiv (x^\alpha)_{\alpha=0}^{d-1}$ denote the first $d$
space-time coordinates, the assumption is that $\Sigma$ can be
parametrized as follows:
\begin{align}\label{eq:defmonge}
\Sigma \big) \; x_\shortparallel \;\to\; y \equiv \big(x_\shortparallel ,
\psi(x_\shortparallel) \big)\;\;, \nonumber\\
	{\rm or:} \; 
y^\mu \,=\, \delta^\mu_\alpha x^\alpha \,+\, \delta^\mu_d \, 
\psi(x_\shortparallel) \;, 
\end{align}
namely, the surface can be described by specifying its height
$\psi(x_\shortparallel)$ at each point $x_\shortparallel$ on the $x^d=0$
hyperplane.

A convenient way to proceed with the evaluation of $\Gamma(\Sigma)$ is by
first representing the functional $\delta$ function by a (functional)
Fourier transform. Indeed, introducing an auxiliary field
$\lambda(x_\shortparallel)$, we may write:
\begin{equation}
\delta_\Sigma(\varphi) \;=\; \int {\mathcal D}\lambda \, e^{i \int
d^dx_\shortparallel \sqrt{g(x_\shortparallel)} \, \lambda(x_\shortparallel)\,
\varphi(x_\shortparallel, \psi(x_\shortparallel)) } \;,
\end{equation}
where $g(x_\shortparallel)$ denotes the determinant of
$g_{\alpha\beta}(x_\shortparallel)$, the induced metric on $\Sigma$:
\begin{equation}\label{eq:metric}
g_{\alpha \beta}(x_\shortparallel) \,=\, \delta_{\alpha \beta} +
\partial_\alpha\psi(x_\shortparallel) \, \partial_\beta
\psi(x_\shortparallel) \;\;,\;\;\;
g(x_\shortparallel) \,= \,  1 + \partial_\alpha\psi(x_\shortparallel) \,
\partial_\alpha \psi(x_\shortparallel)   \;.
\end{equation}
Having in mind, in what follows, this kind of surface and parametrization
for $\Sigma$, we use the self-explanatory notation $\Gamma(\psi)$ for the
effective action. We see that:
\begin{equation}\label{eq:eff_1}
e^{-\Gamma(\psi)}\,=\, \frac{1}{\mathcal N} \, 
\int \mathcal{D}\varphi \mathcal{D}\lambda \; 
	e^{-S_0(\varphi) + i \int_{x_\shortparallel}
\lambda(x_\shortparallel)
\sqrt{g(x_\shortparallel)}\varphi(x_\shortparallel,\psi(x_\shortparallel))}
\end{equation}
where we have used a self-explanatory shorthand notation (that we shall
use in the remainder of the paper) for the integrals: 
\begin{equation}
\int d^{d+1}x \ldots  \equiv \int_x \ldots \;\;,
\int d^dx_\shortparallel \ldots  \equiv \int_{x_\shortparallel} 
\ldots \;\;, \;\; \ldots
\end{equation}

As a first step towards the evaluation of $\Gamma(\psi)$, we see that we
can get rid of the $\sqrt{g(x_\shortparallel)}$ factor in the integral, by
a redefinition of the $\lambda$ field integration measure; namely, 
$\lambda(x_{\shortparallel}) \rightarrow \lambda(x_{\shortparallel})
[g(x_{\shortparallel})]^{-1/2}$,
which yields:
\begin{equation}\label{eq:eff_2}
	e^{-\Gamma(\psi)} \,=\,  \frac{{\mathcal J}_g}{\mathcal N} \,
\int \mathcal{D}\varphi \mathcal{D}\lambda \; 
e^{-S_0(\varphi) + i \int_x J_\Sigma(x) \varphi(x)}
\end{equation}
where ${\mathcal J}_g$ denotes the Jacobian:
\begin{equation}\label{eq:defjacobian}
{\mathcal J}_g \;=\; \det [
		\delta^d(x_\shortparallel-x'_\shortparallel)
		g^{-1/2}(x_\shortparallel) ]
\end{equation}
and:
\begin{equation}\label{eq:J_def}
J_\Sigma(x) \,\equiv \, \lambda(x_\shortparallel) \, 
\delta\left(x_d - \psi(x_\shortparallel)\right) \;.
\end{equation}
Thus, by performing the functional integration of $\varphi$, we obtain for
$\Gamma(\psi)$:
\begin{align}\label{eq:eff_3}
e^{-\Gamma(\psi)} & =\; \det
[\delta^d(x_\shortparallel-x'_\shortparallel)
g^{-1/2}(x_\shortparallel)] \nonumber\\
&\times  \int \mathcal{D}\lambda \; \exp\Big[-\frac{1}{2}
\int_{x_\shortparallel,x'_\shortparallel} \; \lambda(x_{\shortparallel}) \, 
\Delta(x_\shortparallel,x'_\shortparallel) \, \lambda(x'_\shortparallel) 
\Big] \;,
\end{align}
with
\begin{equation}\label{eq:reduced}
\Delta(x_\shortparallel,x'_\shortparallel) \;=\;
\langle x_\shortparallel, \psi(x_\shortparallel) |
(-\partial^2)^{-1} | x'_\shortparallel, \psi(x'_\shortparallel) \rangle \;,
\end{equation}
where 
\begin{equation}
\langle x_\shortparallel, x_d | (-\partial^2)^{-1} |
x'_\shortparallel, x'_d\rangle \;=\; 
\langle x | (-\partial^2)^{-1} | x' \rangle \;=\; 
	\int \frac{d^{d+1}k}{(2\pi)^{d+1}} \frac{e^{i k \cdot (x -
	x')}}{k^2} \;,
\end{equation}
denotes the free scalar field propagator. By integration out $k_d$, we may write
for $\Delta$ a more explicit expression 
\begin{equation}\label{eq:Deltaexp}
\Delta(x_\shortparallel,x'_\shortparallel) \;=\; 
\int \frac{d^dk_{\shortparallel}}{(2\pi)^d} \,
e^{i k_{\shortparallel} \cdot (x_{\shortparallel}-x'_{\shortparallel})}
\, \frac{e^{- |k_{\shortparallel}| |\psi(x_{\shortparallel}) -
\psi(x'_{\shortparallel})|}}{2 |k_{\shortparallel}|} \;.
\end{equation}

The integral over $\lambda$ is a Gaussian. Performing it, produces the
result that allows us to write for $\Gamma(\psi)$ the following:   
\begin{equation}\label{eq:eff_3.1}
e^{-\Gamma(\psi)} \;=\; 
\Big\{ \det \big[ g(x_{\shortparallel}) \, 
\delta^{d}(x_\shortparallel - x'_\shortparallel)\big] \times
\det \big[ \Delta(x_{\shortparallel},x'_{\shortparallel})\big]
\Big\}^{-1/2} \;. 
\end{equation}
Therefore, 
\begin{equation}\label{eq:eff_4}
\Gamma(\psi) \;=\; \Gamma_g(\psi) \,+\, 
\Gamma_\Delta(\psi)
\end{equation}
where $\Gamma_g(\psi) =\frac{1}{2} {\rm Tr}\big\{\log\big[ g(x_\shortparallel) 
\delta^d(x_{\shortparallel}-x'_{\shortparallel}) \big] \big\}$ and:
\begin{equation}
\Gamma_\Delta(\psi) \;=\;
\frac{1}{2} {\rm Tr}\big\{\log\big[\Delta(x_{\shortparallel},x'_{\shortparallel})
\big]\Big\} \;.
\end{equation}

Before proceeding to the perturbative evaluation, we note that the first term, $\Gamma_g(\psi)$, cannot contribute to the imaginary part. Indeed, we first note that it requires the introduction of an $UV$
cutoff $\Lambda$, since it involves a Dirac $\delta$ function evaluated at the
coincident points $x_\shortparallel$, $x'_\shortparallel$:
\begin{equation}\label{eq:Deltag}
\Gamma_g(\psi) \,=\, \frac{1}{2} \Lambda^d \int_{x_\shortparallel}
\log\big[ g(x_\shortparallel)\big] \;.
\end{equation}
In real time, then one has:
\begin{equation}\label{eq:Deltag}
\Gamma_g(\psi) \,=\, \frac{1}{2} \Lambda^d \int_{x_\shortparallel}
\log\big[ 1 - (\partial_0 \psi)^2 + (\nabla_{\shortparallel}
\psi)^2 \big] \;,
\end{equation}
which can only develop an imaginary part for superluminal excitations of
the surface.
Therefore, in what follows we concentrate on the remaining term,
$\Gamma_\Delta$.

\section{Expansion in powers of $\psi$}\label{sec:pert}
With the aim to evaluate $\Gamma_\Delta$, we expand it in powers of $\psi$. To that end, we first expand $\Delta$
\begin{equation}
\Delta \;=\; \Delta^{(0)} \,+\, \Delta^{(1)} \,+\, \ldots 
\end{equation}
where the index denotes order in $\psi$. 
We find:
\begin{align}\label{eq:Deltak}
\Delta^{(l)}(x_{\shortparallel},x'_{\shortparallel})  & = \; 
\int_{\not k_\shortparallel} \, 
\frac{e^{ik_{\shortparallel}(x_{\shortparallel}-x'_{\shortparallel})}}{2\big
|k_{\shortparallel}\big|} \;  \frac{(-1)^l}{l!} \big|
k_{\shortparallel}\big|^l \big|  \psi(x_{\shortparallel}) - 
\psi(x'_{\shortparallel} ) \big|^l \;, 
\end{align}
where we introduced the shorthand notation
\begin{equation}
\int_{\not \! k_\shortparallel} \ldots \; \equiv\; 
\int \frac{d^dk}{(2\pi)^d} \ldots 
\end{equation}

We note that $\Delta^{(k)}$ vanishes for odd $k$. To
prove that, it is convenient to undo the integral over $k_d$, which lead to
(\ref{eq:Deltaexp}). 
Then one realizes that
\begin{align}
\Delta^{(k)}(x_\shortparallel, x'_\shortparallel ) 
& = \; \frac{i^k}{k!} \, \int \frac{d^d p_\shortparallel}{(2\pi)^d} \,
e^{i p_\shortparallel \cdot (x_\shortparallel - x'_\shortparallel)}
\int_{-\infty}^{+\infty}\frac{dp_d}{2\pi} 
\frac{p_d^k}{p_\shortparallel^2 + p_d^2} 
\big( \psi(x_\shortparallel) - \psi(x'_\shortparallel) \big)^k 
\nonumber\\
& = \, 0 \;\;,\;\;{\rm for}\, {\rm any} \, {\rm odd} \, k \;.
\end{align}

Hence the original expression for $\Delta$ may in fact be represented equivalently
as follows:
\begin{equation}
\Delta(x_\shortparallel, x'_\shortparallel ) 
	\; = \; \int_{\not p_\shortparallel} \,
e^{i p_\shortparallel \cdot (x_\shortparallel - x'_\shortparallel)}
\frac{ \cosh\big[|p_\shortparallel| (\psi(x_\shortparallel) -
\psi(x'_\shortparallel))\big]}{2 |p_\shortparallel| } \;.
\end{equation}

Using the expansion of the $\cosh$, 
\begin{equation}\label{eq:eff_6}
\Gamma_\Delta(\psi) \;=\; \frac{1}{2} {\rm Tr}\Big[ \log\big(
\Delta^{(0)} \big)\Big]  +  \frac{1}{2} {\rm Tr}\Big[ \log\big( 1 +
\sum\limits_{k=1} A^{(2 k)} \big)\Big]   
\end{equation}
where $A^{(2 k)}$ stands for
\begin{equation}\label{eq:Al}
A^{(2 k)}(x_{\shortparallel},x'_{\shortparallel}) \;=\; 
\int_{y_\shortparallel} \; 
\big[\Delta^{(0)}\big]^{-1}(x_{\shortparallel},y_{\shortparallel})
\; \Delta^{(2 k)}(y_{\shortparallel},x'_{\shortparallel}) \;.
\end{equation}

We shall discard the first, zeroth order term in the expansion for
$\Gamma_\Delta$, since it is a constant independent of
$\psi$, and it represents the (divergent) effective action corresponding to the
infinite plane $x_d = 0$. In the context of this expansion, it is the
same for any surface. Thus, in what follows, 
$\Gamma_\Delta(\psi) \equiv \frac{1}{2} {\rm Tr}\Big[ \log\big( 1 +
\sum\limits_{k=1} A^{(2 k)} \big)\Big]$, which upon expansion of the logarithm yields,
\begin{equation}\label{eq:logexp}
\log\left( 1 + \sum_{k=1}^\infty A^{(2 k)} \right) \,=\, 
\sum_{n=1}^\infty \frac{(-1)^{n-1}}{n} 
	\big[\sum_{k=1}^\infty A^{( 2 k)}\big]^n \;,
\end{equation}
leading to   
\begin{align} 
\Gamma_\Delta(\psi) \;=\; \frac{1}{2} \, \sum_{k=1}^\infty {\rm Tr}\big[A^{(2 k)}\big]
\,-\, \frac{1}{4} \sum_{k=2}^\infty \sum_{l=1}^{k-1} {\rm Tr}\big[
A^{(2 k - 2 l)} \,  A^{(2 l)}\big] \nonumber\\
+ \; \frac{1}{6} \, 
\sum_{k=3}^\infty \sum_{l=2}^{k-1} \sum_{m=1}^{l-1}{\rm Tr}\big[
A^{(2 k - 2 l)} \, A^{(2 l - 2 m)} \,  A^{(2 m)}\big] \;+\; \dots
\end{align}

  Regrouping terms,
\begin{equation}
\Gamma_\Delta(\psi) \;=\; \Gamma_\Delta^{(2)}(\psi) \,+\,
\Gamma_\Delta^{(4)}(\psi) \,+\, \Gamma_\Delta^{(6)}(\psi) \,+\, \ldots 
\end{equation}
where the explicit form of the first few terms is:
\begin{align} \label{Gamma all orders}
\Gamma_\Delta^{(2)}(\psi) &=\;  \frac{1}{2} {\rm Tr}\big[A^{(2)}\big] \;,  
\nonumber\\ 
\Gamma_\Delta^{(4)}(\psi) &=\;  \frac{1}{2} {\rm Tr}\big[A^{(4)}\big]   
\,-\,\frac{1}{4} {\rm Tr}\big[A^{(2)} A^{(2)}\big] \nonumber\\
\Gamma_\Delta^{(6)}(\psi) &=\;  \frac{1}{2} {\rm Tr}\big[A^{(6)}\big]   
\,-\,\frac{1}{2} {\rm Tr}\big[A^{(4)} A^{(2)}\big]
	\,+\,\frac{1}{6} {\rm Tr}\big[A^{(2)} A^{(2)} A^{(2)} \big]
 \nonumber\\
\ldots
\end{align}
Let us now evaluate the first two terms, namely, the ones of order 2 and 4 in $\psi$.
\subsection{$\Gamma_\Delta^{(2)}$}
\begin{figure}[h!] 
	\centering
	\includegraphics[width=0.5\textwidth]{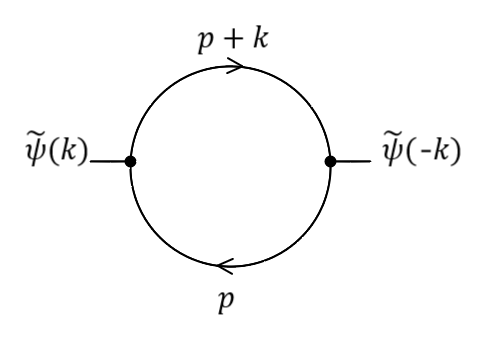}
	\caption{Diagramatic representation of the second order contribution $\Gamma^{(2)}_\Delta$.}
	\label{fig:diagram order 2}
\end{figure}

Introducing the explicit form of $A^{(2)}$, we see that:
\begin{align} \label{Development k=2 n=1}
   \Gamma_\Delta^{(2)}&= \frac{1}{4} \int \limits_{x\, y \,\not{k} \; \not{k}' } \, e^{ik(x-y)} \, e^{ik'(y-x)} \; \left |  k\right |\left |  k'\right | \, \left( \psi^2(x) + \psi^2(y) - 2\psi(x) \psi(y)   \right) \\ \nonumber
   &= \frac{1}{2} \int \limits_{x\,\not{k} \; \not{k}'} (2\pi)^d \delta^d(k-k') \, e^{ix(k-k')}|\left | \psi(x) \;\right |^2 \left |  k\right |\left |  k'\right | \\ \nonumber 
   & - \frac{1}{2} \int \limits_{x \,y\,\not{k} \; \not{k}'} \, e^{ix(k-k')} \;\psi(x) \;e^{iy(k'-k)}\;\psi(y) \left |  k\right |\left |  k'\right | \\ \nonumber
   &=\frac{1}{2} \int\limits_{\cancel{k}\, \cancel{p}} \,\left |\tilde{\psi}(k) \right |^2\,  \left |  p\right |^2 - \frac{1}{2} \int \limits_{\cancel{k} \, \cancel{p}} \; \left |  p\right |\left |  p+k\right | \; \left | \tilde{\psi}(k) \right |^2 .
\end{align}
So we can write
\begin{equation}\label{eq:gamma_2}
\Gamma_\Delta^{(2)}\;=\; \frac{1}{2} \int_{\not k_\shortparallel} \, 
\big[ F(0) - F(k_\shortparallel) \big] \,
\big|\widetilde{\psi}(k_\shortparallel)\big|^2\, , 
\end{equation}
where the tilde denotes Fourier transformation ($k_\shortparallel$ is a $d$
component momentum). We now evaluate the integral over $p_\shortparallel$,
rendering $F$ in a form that emphasizes the fact that it is, indeed, a loop integral, albeit with propagators having non-standard exponents (see Fig. \ref{fig:diagram order 2} ).

Namely, 
\begin{equation}\label{F_definition}
F(k_\shortparallel) \,=\, 
\int_{\not p_\shortparallel} \frac{1}{(p_\shortparallel^2)^{-\frac{1}{2}}
[(p_\shortparallel+k_\shortparallel)^2]^{-\frac{1}{2}}} \;.
\end{equation}
In Fig. \ref{fig:diagram order 2}  a Feynman diagram depicts this
contribution. An internal line with momentum $p$ represents
$|p_\shortparallel|$. The argument of $\widetilde{\psi}$
is positive for ingoing momenta. 

The loop integral in (\ref{F_definition}) is superficially divergent. The (required) zero-momentum subtraction is insufficient to render it finite, but
it is possible, however, to apply a procedure devised in a previous reference \cite{Fosco:2007nz}
, involving an analytic regularization. This involves a continuation of the exponents in the would-be propagators, as well as of
the number of dimensions. After evaluating the momentum integral, the
continuation back to the physical values provides results which allow us to obtain the required result for the imaginary parts. 

We proceed then to introduce a Feynman parameter $\alpha$,
to generate an expression which generalizes the loop integral we actually need:
\begin{equation}\label{F Feynman parameters}
F_{\lambda_1,\lambda_2}(k_\shortparallel)  
 \;=\; \frac{\Gamma(\lambda_1 + \lambda_2)}{\Gamma(\lambda_{1})\Gamma(\lambda_2)} 
\, \int_0^1 d\alpha \;\int _{\not p_\shortparallel} 
	\frac{\alpha^{\lambda_1-1}(1-\alpha)^{\lambda_2-1}}{[\alpha~p_\shortparallel^2+(1-\alpha)
(p_\shortparallel+k_\shortparallel)^2]^{\lambda_1+\lambda_2}}
\end{equation}
so that $F= F_{\lambda_1,\lambda_2}\big|_{\lambda_1, \lambda_2 \to - \frac{1}{2}}$ . 

Integrating out the loop momentum,
\begin{equation}\label{F Loop integration}
F_{\lambda_1, \lambda_2}(k_\shortparallel) \,=\,
\frac{\Gamma(\lambda_1 + \lambda_2 - d/2)}{(4\pi)^{\frac{d}{2}} \Gamma(\lambda_{1})\Gamma(\lambda_2)} \,\int_{0}^{1} d\alpha \; \alpha^{\frac{d}{2} - \lambda_2 -1 }(1-\alpha)^{\frac{d}{2} -\lambda_1 -1}  \,
(k_\shortparallel^2)^{\frac{d}{2} -\lambda_1-\lambda_2} 
\end{equation}
so that, for any $d$:
\begin{align}\label{eq:F2}
	F(k_\shortparallel)&=  \frac{1}{(4\pi)^{\frac{d}{2}}} \,
	\frac{\Gamma(- 1 - \frac{d}{2})}{[\Gamma(-\frac{1}{2})]^2} 
	\,\int_{0}^{1} d\alpha \; \big[ \alpha  
	(1-\alpha) \big]^{\frac{d - 1}{2}} 
\; (k_\shortparallel^2)^{\frac{d+2}{2}} \;,
\end{align}
or, after integrating $\alpha$,
\begin{equation}\label{eq:F3}
	F(k_\shortparallel) \,=\,  \frac{1}{(4\pi)^{\frac{d}{2}}} \,
	\frac{\Gamma(- 1 - \frac{d}{2}) [\Gamma(\frac{d + 1}{2}) ]^2}{[\Gamma(-\frac{1}{2})]^2 \Gamma(d + 1) } \, (k_\shortparallel^2)^{\frac{d+2}{2}}  \;,
\end{equation}
which is a general expression for the kernel of the second order contribution, in principle valid for any $d$, which we recall is the number of spatial dimensions. Some remarks are in order here before obtaining more explicit expressions in particular cases. Firstly, we note that no subtraction at zero momentum is necessary, as the dimensional regularization procedure has already gotten rid of the divergences, subtracting the potentially divergent terms.

To proceed, we note that it is important to distinguish between odd and even values of $d$.
\subsubsection{Odd number of spatial dimensions  $d = 2 q +1$ ($q = 0, 1, \ldots )$} \label{Order 2 odd d}
In this case, there is no divergence when taking the limit $d \to (2 q +1)$. 
We find:
\begin{equation}\label{eq:Fodd}
	F(k_\shortparallel) \,=\,  \frac{(-1)^q}{(2\pi)^{q+1}}\,
	\frac{(q!)^2}{(2 q + 1)! (2 q + 3)!!} \; (k_\shortparallel^2)^{q + \frac{3}{2}}  \;.
\end{equation}

Therefore,
\begin{equation}\label{eq:gamma_odd_e}
    \Gamma^{(2)}_\Delta \; = \;
    \frac{(-1)^{q+1} (q!)^2}{2 (2\pi)^{q+1} (2 q + 1)! (2 q + 3)!!} \; 
     \int_{\not k_\shortparallel} |\widetilde{\psi}(k_\shortparallel)|^2 \,  (k_\shortparallel^2)^{q + 1 + \frac{1}{2}}  \;.
\end{equation}

Let us obtain here the real-time version of this contribution, obtained by continuation of the time component of the momentum. We note that the Euclidean $k_0$ should be replaced by $-i$ times its real time counterpart, and that $(-)$ times the Euclidean effective action becomes $i$ times its real time version.
Denoting by $\Gamma^{(2)}$ the real time effective action corresponding to    $\Gamma^{(2)}_\Delta$, we get:
\begin{equation}\label{eq:gamma_odd_r}
	\Gamma^{(2)}\; = \;
	\frac{(-1)^{q+1} (q!)^2}{2 (2\pi)^{q+1} (2 q + 1)! (2 q + 3)!!} \; 
	\int_{\not k_\shortparallel}  \,  [{\mathbf k}_\shortparallel^2 - (k^0)^2 ]^{q + 1 + \frac{1}{2}} \;  |\widetilde{\psi}(k^0 , 
	{\mathbf k_\shortparallel})|^2\;,
\end{equation}
where, in a natural notation, we have set $k_\shortparallel = (k^0 , 
{\mathbf k_\shortparallel})$.

We then see that the threshold condition for the appearance of an imaginary part in $\Gamma^{(2)}$ is that the Fourier transform of $\psi$ has components with  $|k^0| > |{\mathbf k}_\shortparallel |$.  Its explicit form is:
\begin{equation}\label{eq:im_gamma_odd_r}
 	{\rm Im} \big[ \Gamma^{(2)} \big] \; = \;
 	\frac{1}{2} \; \eta_{2q+1} \, \int_{\not k_\shortparallel} \theta(|k^0| - |{\mathbf k}_\shortparallel|) \, \,  [(k^0)^2 - {\mathbf k}_\shortparallel^2 ]^{q + 1 + \frac{1}{2}} 
 	 \; |\widetilde{\psi}(k^0 , 
 	{\mathbf k_\shortparallel})|^2\;,
\end{equation}
with $\eta_{2q+1} = \frac{(q!)^2}{(2\pi)^{q+1} (2 q + 1)! (2 q + 3)!!}$. Note that $\eta_{2q+1}$ is positive, which complies with the fact that the probability of vacuum decay must be smaller than $1$.
\subsubsection{Even number of spatial dimensions  $d = 2 q$ ($q = 1, 2, \ldots )$} \label{Order 2 even d}
Since $\Gamma(- 1 - \frac{d}{2})$ has a pole for even $d$, we set $d = (2 q - \epsilon)$, and consider the limit when $\epsilon \to 0$. Note that, to keep the mass dimensions correct, we need to introduce a mass parameter $\mu$.
We find, to first order in  $\epsilon$:
\begin{equation}\label{eq:Feven}
F(k_\shortparallel) \,=\,  \frac{(-1)^{q+1}}{(4\pi)^q}\,
    	\frac{\big[\Gamma(q + \frac{1}{2}) \big]^2}{(q + 1 )! (2 q)! \big[\Gamma( -\frac{1}{2}) \big]^2} \; 
    	\left(\frac{2}{\epsilon} - \gamma_E\right)  (k_\shortparallel^2)^{q + 1}
    	\big[ 1 - \frac{\epsilon}{2}  \log (\frac{k_\shortparallel^2}{4 \pi \mu^2}) \big] 
    	  \;.
\end{equation}
where $\gamma_E$ is the Euler-Mascheroni constant.

One of the contributions is divergent when $\epsilon \to 0$, and corresponds to a counterterm which is analytic in $k_\shortparallel^2$, and thus cannot contribute to the imaginary part, while the term proportional to $\gamma_E$ is also analytic in $k_\shortparallel^2$. This leaves just one contributing term, since it involves a $\log$ function that can have a negative argument, 
 \begin{equation}\label{eq:Feven_1}
 F(k_\shortparallel) \,=\,  \frac{(-1)^{q+2}}{(4\pi)^q}\,
 \frac{\big[\Gamma(q + \frac{1}{2}) \big]^2}{(q + 1 )! (2 q)! \big[\Gamma( -\frac{1}{2}) \big]^2} \;  (k_\shortparallel^2)^{q + 1} \log (\frac{k_\shortparallel^2}{4 \pi \mu^2}) 
 \; + \,\textrm{real terms}.
 \end{equation}   
Therefore,
\begin{equation}\label{eq:im_gamma_even_r}
{\rm Im} \big[ \Gamma^{(2)} \big] \; = \;
\frac{1}{2} \; \eta_{2q} \, \int_{\not k_\shortparallel} \theta(|k^0| - |{\mathbf k}_\shortparallel|) \, \,  [(k^0)^2 - {\mathbf k}_\shortparallel^2 ]^{q + 1} 
\; |\widetilde{\psi}(k^0 , {\mathbf k_\shortparallel})|^2\;,
\end{equation}   
where now $\eta_{2q} = \frac{ \pi}{(4\pi)^q}\,
\frac{\big[\Gamma(q + \frac{1}{2}) \big]^2}{(q + 1 )! (2 q)! \big[\Gamma( -\frac{1}{2}) \big]^2}$. 
    
\begin{figure}[h] 
     \centering
     \includegraphics[width=0.7\textwidth]{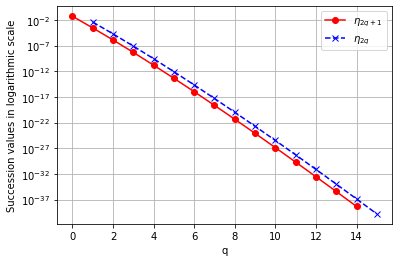}
     \caption{Successions $\eta_{2q+1}$ and $\eta_{2q}$ in logarithmic scale.}
     \label{fig:ratio_order2}
\end{figure}

 The successions $\eta_{2q+1}$ and $\eta_{2q}$ behave similarly with growing dimensions, since they are both positive and decrease rapidly as functions of $q$. We plot them in logarithmic scale in Fig. \ref{fig:ratio_order2}, where we can see their linear behaviour with negative slopes, which indicates an exponential decline in their values.

\subsection{Evaluation of $\Gamma^{(4)}_\Delta$}
Because of the complexity of the full expression for this term, we shall focus here on its evaluation for a particular, but rather relevant excitation of the surface: a plane wave characterized by a wave vector $k_\shortparallel$.
As we shall see,  the appearance of higher powers of $\tilde{\psi}$, due to the higher powers of $A$ in \eqref{Gamma all orders},  will translate into higher multiples of the momentum vector $k_\shortparallel$ in the decay probability. 

For the sake of clarity, we will divide this contribution as $\Gamma_\Delta^{(4)}=\Gamma_\Delta^{(4,1)}+\Gamma_\Delta^{(4,2)}$, where
\begin{equation} \label{Equation k=4 n=1}
    \Gamma_\Delta^{(4,1)}= \frac{1}{2}     {\rm Tr }\left[A^{(4)}   \right].
\end{equation}
and
\begin{equation} \label{Equation k=2 n=2}
    \Gamma_\Delta^{(4,2)}=- \frac{1}{4}     {\rm Tr }\left[A^{(2)}\,A^{(2)}   \right] \;.
\end{equation}

So we begin with:
\begin{align} \label{Development k=4 n=1}
    \Gamma_\Delta^{(4,1)}&= \frac{1}{48}\int_{x_\shortparallel \, x'_\shortparallel \, \not{k_\shortparallel} \, \not{k_\shortparallel}'} e^{ik_\shortparallel(x_\shortparallel-x'_\shortparallel)} e^{ik_\shortparallel'(x_\shortparallel'-x_\shortparallel)} \left| k_\shortparallel\right| \left| k'_\shortparallel\right|^3 \left( \psi(x_\shortparallel) - \psi(x'_\shortparallel) \right)^4 \\ \nonumber
    &=\frac{1}{24} \int_{x_\shortparallel \, \not{k_\shortparallel}} \,\left|k_\shortparallel\right|^4 \psi(x_\shortparallel)^4 + \frac{1}{8} \int \limits_{\not{k_\shortparallel}} \tilde{\psi^2}(k_\shortparallel) \tilde{\psi^2}(-k_\shortparallel)\; G(k_\shortparallel) - \frac{1}{6} \int_{\not{k_\shortparallel}} \tilde{\psi^3}(k_\shortparallel) \tilde{\psi}(-k_\shortparallel) \;G(k_\shortparallel) \\ \nonumber
    &= \int_{\not{k_\shortparallel}} \left[ \tilde{\psi^2}(k_\shortparallel) \tilde{\psi^2}(-k_\shortparallel) \left( \frac{G(0_\shortparallel)}{24} + \frac{G(k_\shortparallel)}{8} \right) - \frac{1}{6} \tilde{\psi^3}(k_\shortparallel) \tilde{\psi}(-k_\shortparallel) \;G(k_\shortparallel)  \right],
\end{align}
where
\begin{equation} \label{G definition}
   G(k_\shortparallel)=\int _{\not{p_\shortparallel}} \left| p_\shortparallel \right|^3 \left| p_\shortparallel + k_\shortparallel \right| = \int _{\not{k}} \frac{1}{(p_\shortparallel^2)^{-\frac{3}{2}}} \frac{1}{((p_\shortparallel+k_\shortparallel)^2)^{-\frac{1}{2}}} \;.
\end{equation}
The different terms in this contribution are illustrated in Fig. \ref{fig: Order(4,1)}. The number of legs of each external line indicates the power of $\psi$, and for internal lines, their number indicates the power of the momentum they carry. Note that the total number of external legs should always match the total number of internal lines.

\begin{figure}[h] 
     \centering
     \includegraphics[width=0.7\textwidth]{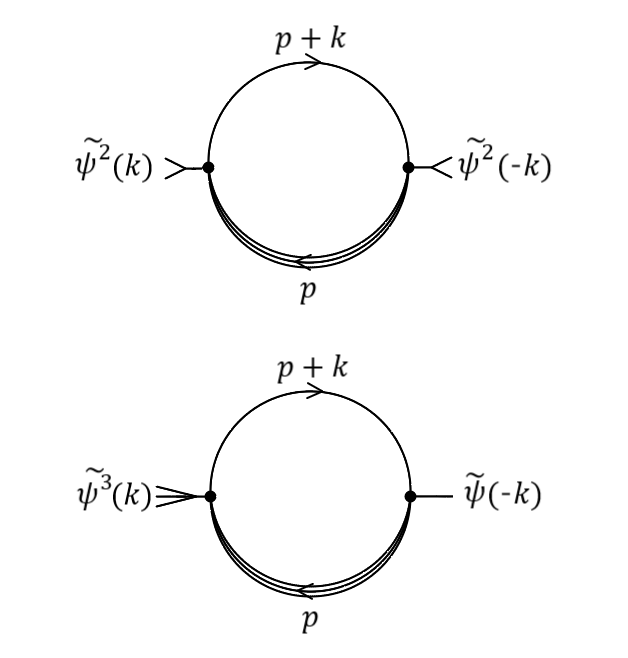}
     \caption{Diagramatic representation of the fourth order contribution 
$\Gamma_\Delta^{(4,1)}$.}
     \label{fig: Order(4,1)}
\end{figure}

Using essentially the same proccedure as in \eqref{F Feynman parameters}, we get:
\begin{align} \label{G loop integration}
    G(k_\shortparallel)&=  \frac{\Gamma(\lambda_{1}+\lambda_{2}-\frac{d}{2})}{\Gamma(\lambda_{1})\Gamma(\lambda_2)} \int\limits_{0}^{1}d\alpha \; \alpha^{\lambda_1-1}(1-\alpha)^{\lambda_2-1} \frac{\left( \alpha(1-\alpha) k_\shortparallel^2 \right)^{\frac{d}{2}-(\lambda_1+\lambda_2)}}{(4\pi)^{\frac{d}{2}}} \\ \nonumber
&= \frac{\Gamma(\lambda_{1}+\lambda_{2}-\frac{d}{2})}{(4\pi)^{\frac{d}{2}} \Gamma(\lambda_1) \Gamma(\lambda_2)}  \int \limits_0^1 d\alpha \;\alpha^{\frac{d}{2} - \lambda_2 -1 }(1-\alpha)^{\frac{d}{2} -\lambda_1 -1}  \,
(k_\shortparallel^2)^{\frac{d}{2} -\lambda_1-\lambda_2} \\ \nonumber
&=\frac{\Gamma(-2-\frac{d}{2}) \, \Gamma(\frac{d+3}{2}) \Gamma(\frac{d+1}{2})}{(4\pi)^{\frac{d}{2}} \Gamma(-\frac{3}{2}) \Gamma(-\frac{1}{2}) \Gamma(d+2) } (k_\shortparallel^2)^{\frac{d}{2}+2} \,.
\end{align}

Now we move on to the second term of the contribution,

\begin{align} \label{Development 1 k=2 n=2}
     \Gamma_\Delta^{(4,2)}&=- \frac{1}{4}     {\rm Tr }\left[A^{(2)}\,A^{(2)}   \right] \\ \nonumber
     &=-\frac{1}{16} \int_{X_\shortparallel,\not{K_\shortparallel}}e^{i\left[ k_\shortparallel(x_\shortparallel-y_\shortparallel) + k'_\shortparallel(y_\shortparallel-y'_\shortparallel) + q_\shortparallel(y'_\shortparallel-x'_\shortparallel) + q'_\shortparallel(x'_\shortparallel-x_\shortparallel) \right]} \\ \nonumber &\times\left|k_\shortparallel\right|\left|k'_\shortparallel\right|\left|q_\shortparallel\right|\left|q'_\shortparallel\right| \left(\psi(y_\shortparallel)-\psi(y'_\shortparallel)\right)^2 \left(\psi(x_\shortparallel)-\psi(x'_\shortparallel)\right)^2
\end{align}
where $ \int_{X_\shortparallel,\not{K_\shortparallel}}$ indicates integration over all space-time and momentum variables respectively. This integral separates into 9 terms, since

\begin{align}\label{Development 2 k=2 n=2}
&\left(\psi(x_\shortparallel)-\psi(x'_\shortparallel)\right)^2 \left(\psi(y_\shortparallel)-\psi(y'_\shortparallel)\right)^2 =\\ \nonumber
&= \psi^2(x_\shortparallel) \psi^2(y_\shortparallel) +\psi^2(x_\shortparallel) \psi^2(y'_\shortparallel) +\psi^2(x'_\shortparallel) \psi^2(y_\shortparallel) + \psi^2(x'_\shortparallel) \psi^2(y'_\shortparallel) \\ \nonumber 
&-2\left[\psi(x_\shortparallel) \psi(x'_\shortparallel) \psi^2(y_\shortparallel) + \psi(x_\shortparallel) \psi(x'_\shortparallel) \psi^2(y'_\shortparallel) + \psi(y_\shortparallel) \psi(y'_\shortparallel) \psi^2(x_\shortparallel) + \psi(y_\shortparallel) \psi(y'_\shortparallel) \psi^2(x'_\shortparallel)\right] \\ \nonumber
&+4 \psi(x_\shortparallel) \psi(x'_\shortparallel) \psi(y_\shortparallel) \psi(y'_\shortparallel).
\end{align}

 For convenience, we will separate \eqref{Development 2 k=2 n=2} into three different contributions that have different kernels, as we will show below. We write said separation as

\begin{equation} \label{Development 3 k=2 n=2}
    \Gamma_\Delta^{(4,2)}=-\frac{1}{16} \Gamma_A + \frac{1}{8} \Gamma_B - \frac{1}{4} \Gamma_C,
\end{equation}
being $\Gamma_A$ the integral with the four terms in the first line in \eqref{Development 2 k=2 n=2}, $\Gamma_B$ the one with the second line, and $\Gamma_C$ the one with the third line. 

We start with

\begin{align} \label{Development 4 k=2 n=2}
    \Gamma_A &= \int_{X_\shortparallel,\not{K_\shortparallel}} e^{i\left[ k_\shortparallel(x_\shortparallel-y_\shortparallel) + k'_\shortparallel(y_\shortparallel-y'_\shortparallel) + q_\shortparallel(y'_\shortparallel-x'_\shortparallel) + q'_\shortparallel(x'_\shortparallel-x_\shortparallel) \right]}\left|k_\shortparallel\right|\left|k'_\shortparallel\right|\left|q_\shortparallel\right|\left|q'_\shortparallel\right|  \times \\ \nonumber
    &\left[\psi^2(x'_\shortparallel) \psi^2(y'_\shortparallel) + \psi^2(x'_\shortparallel) \psi^2(y_\shortparallel) + \psi^2(x_\shortparallel) \psi^2(y'_\shortparallel) + \psi^2(x_\shortparallel) \psi^2(y_\shortparallel) \right] \\ \nonumber
    &=2\int_{\not{K_\shortparallel}} (2\pi)^{2d} \delta^d(q'_\shortparallel-k_\shortparallel) \delta^d(k_\shortparallel-k'_\shortparallel) \tilde{\psi^2}(q_\shortparallel-q'_\shortparallel) \tilde{\psi^2}(k'_\shortparallel-q_\shortparallel) \left|k_\shortparallel\right|\left|k'_\shortparallel\right|\left|q_\shortparallel\right|\left|q'_\shortparallel\right| \\ \nonumber
    &+2\int_{\not{K_\shortparallel}} (2\pi)^{2d} \delta^d(q'_\shortparallel-q_\shortparallel) \delta^d(k_\shortparallel-k'_\shortparallel) \tilde{\psi^2}(q_\shortparallel-q'_\shortparallel) \tilde{\psi^2}(k'_\shortparallel-q_\shortparallel) \left|k_\shortparallel\right|\left|k'_\shortparallel\right|\left|q_\shortparallel\right|\left|q'_\shortparallel\right| \\ \nonumber
    &=2\int_{\not{q_\shortparallel}\,\not{q_\shortparallel}'} \left(|q'_\shortparallel|^3|q_\shortparallel| + |q'_\shortparallel|^2|q_\shortparallel|^2\right)\, \tilde{\psi^2}(q_\shortparallel-q'_\shortparallel) \tilde{\psi^2}(q'_\shortparallel-q_\shortparallel) \\ \nonumber
    &=2\int _{\not{k_\shortparallel}} \tilde{\psi^2}(k_\shortparallel) \tilde{\psi^2}(-k_\shortparallel) \; G(k_\shortparallel) + \,\text{real term}\,,
\end{align}
with $G(k_\shortparallel)$ defined by \eqref{G loop integration}. This corresponds to the first diagram in Fig. \ref{fig: Order(4,1)}. We're not interested in the term with the product $|q'_\shortparallel|^2|q_\shortparallel|^2$ since it can't contribute to the imaginary part, so we discard it. Then we procceed with the term

\begin{align} \label{Development 5 k=2 n=2}
    \Gamma_B&=\int_{X_\shortparallel,\not{K_\shortparallel}}e^{i\left[ k_\shortparallel(x_\shortparallel-y_\shortparallel) + k'_\shortparallel(y_\shortparallel-y'_\shortparallel) + q_\shortparallel(y'_\shortparallel-x'_\shortparallel) + q'_\shortparallel(x'_\shortparallel-x_\shortparallel) \right]} \left|k_\shortparallel\right|\left|k'_\shortparallel\right|\left|q_\shortparallel\right|\left|q'_\shortparallel\right| \times \\ \nonumber
    & \left[\psi(x_\shortparallel) \psi(x'_\shortparallel) \psi^2(y_\shortparallel) + \psi(x_\shortparallel) \psi(x'_\shortparallel) \psi^2(y'_\shortparallel) + \psi(y_\shortparallel) \psi(y'_\shortparallel) \psi^2(x_\shortparallel) + \psi(y_\shortparallel) \psi(y'_\shortparallel) \psi^2(x'_\shortparallel)\right]  \\ \nonumber
    &=2\int_{\not{p_\shortparallel}\, \not{q_\shortparallel}} \left( \tilde{\psi}(-p_\shortparallel) \tilde{\psi}(-q_\shortparallel) \tilde{\psi^2}(p_\shortparallel+q_\shortparallel) H(p_\shortparallel,q_\shortparallel) + \tilde{\psi}(p_\shortparallel) \tilde{\psi}(q_\shortparallel) \tilde{\psi^2}(-p_\shortparallel-q_\shortparallel) H(-q_\shortparallel,-p_\shortparallel) \right),
\end{align}
where
\begin{equation} \label{H definition}
H(p_\shortparallel,q_\shortparallel)= \int_{\not{k_\shortparallel}} \left|k_\shortparallel\right|^2\left|k_\shortparallel-p_\shortparallel\right|\left|k_\shortparallel-p_\shortparallel-q_\shortparallel\right|=\int \limits_{\not{k_\shortparallel}} \frac{k_\shortparallel^2}{((k_\shortparallel-p_\shortparallel)^2)^{-\frac{1}{2}} ((k_\shortparallel-p_\shortparallel-q_\shortparallel)^2)^{-\frac{1}{2}}}.    
\end{equation}

\begin{figure}[h] 
     \centering
     \includegraphics[width=0.7\textwidth]{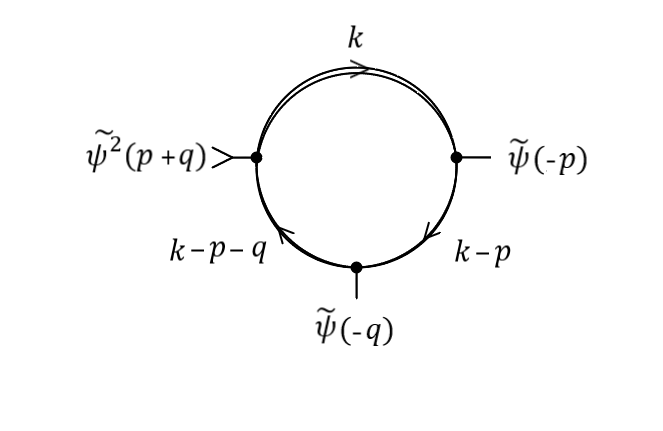}
     \caption{Diagramatic representation of the fourth order contribution 
$\Gamma_B$.}
     \label{fig: Order 4 C}
\end{figure}

This contribution is illustrated in Fig. \ref{fig: Order 4 C}. We can
deal with \eqref{H definition} as in \eqref{F Feynman parameters} and
\eqref{G loop integration},
\begin{align}\label{H Feynman parameters}
&\int_{\not{k_\shortparallel}}
\frac{k_\shortparallel^2}{((k_\shortparallel-p_\shortparallel)^2)^{-\frac{1}{2}}
((k_\shortparallel-p_\shortparallel-q_\shortparallel)^2)^{-\frac{1}{2}}}=
\nonumber\\
& = \; \frac{\Gamma(\lambda_1 + \lambda_2)}{\Gamma(\lambda_{1})\Gamma(\lambda_2)}
\int_0^1 d\alpha \;
\alpha^{\lambda_1-1}(1-\alpha)^{\lambda_2-1} \,
\int_{\not{p_\shortparallel}}
\frac{g_{\mu\nu}\,k_\shortparallel^\mu
	k_\shortparallel^\nu}{(\alpha~p^2+(1-\alpha)(p_\shortparallel+k_\shortparallel)^2)^{\lambda_1+\lambda_2}} \, ,
\end{align}
where we plug in $\lambda_1=\lambda_2=-\frac{1}{2}$ after integration.
Since this equation has components of the loop variable on the numerator,
we use the following formulae:
\begin{equation} \label{Integral formula 1}
    \int _{\not{k_\shortparallel}} \frac{k_\shortparallel^\mu k_\shortparallel^\nu}{\left( k_\shortparallel^2+ 2\,Q\,k_\shortparallel + C\right)^{a} } = \left( Q^\mu Q^\nu + g^{\mu \nu} \frac{(C-Q^2)}{2} \frac{\Gamma(a-\frac{d}{2}-1)}{\Gamma(a-\frac{d}{2})} \right) I_0
\end{equation}
with
\begin{equation}\label{Integral formula 2}
    I_0=\frac{\left(C-Q^2\right)^{\frac{d}{2}-a}}{(4\pi)^{\frac{d}{2}}} \frac{\Gamma(a-\frac{d}{2})}{\Gamma(a)}
\end{equation}
in Euclidean formalism, where $g^{\mu\nu}$ is just the identity.

Then
\begin{align}\label{H loop integration}
& H(p_\shortparallel , q_\shortparallel) \;=\;
\frac{\Gamma(-1-\frac{d}{2})}{\Gamma(-\frac{1}{2})\Gamma(-\frac{1}{2})} 
\frac{(q_\shortparallel^2)^{\frac{d}{2}+1}}{(4\pi)^{\frac{d}{2}}}
	\nonumber\\
&\times \Big[ p_\shortparallel^2
\frac{(\Gamma(\frac{d+1}{2}))^2}{\Gamma(d+1)} + 2 p_\shortparallel q_\shortparallel
\frac{\Gamma(\frac{d+1}{2})\Gamma(\frac{d+3}{2})}{\Gamma(d+2)} +
q_\shortparallel^2
	\frac{\Gamma(\frac{d+1}{2})\Gamma(\frac{d+5}{2})}{\Gamma(d+3)}
\nonumber\\
& + \; \frac{d}{2} q_\shortparallel^2 \,
\frac{(\Gamma(\frac{d+3}{2}))^2}{\Gamma(d+3)}\,
\frac{\Gamma(-2-\frac{d}{2})}{\Gamma(-1-\frac{d}{2})} \Big] \,. 
\end{align}

Finally, 
\begin{equation}\label{eq:dev7}
\Gamma_C \;=\; \int
	_{\not{k_\shortparallel}\,\not{p_\shortparallel}\,\not{q_\shortparallel}}
	\tilde{\psi}(-k_\shortparallel) \tilde{\psi}(-p_\shortparallel)
	\tilde{\psi}(-q_\shortparallel)
	\tilde{\psi}(k_\shortparallel+p_\shortparallel+q_\shortparallel) \,
	K(k_\shortparallel,p_\shortparallel,q_\shortparallel),
\end{equation}
with 
\begin{equation} \label{K definition}
 K(k_\shortparallel,p_\shortparallel,q_\shortparallel)=\int
	_{\not{\ell}_\shortparallel} |\ell_\shortparallel|
	|\ell_\shortparallel+k_\shortparallel|
	|\ell_\shortparallel+k_\shortparallel+p_\shortparallel|
	|\ell_\shortparallel+k_\shortparallel+p_\shortparallel+q_\shortparallel|
	\;.
\end{equation}

\begin{figure}[h] 
     \centering
     \includegraphics[width=0.65\textwidth]{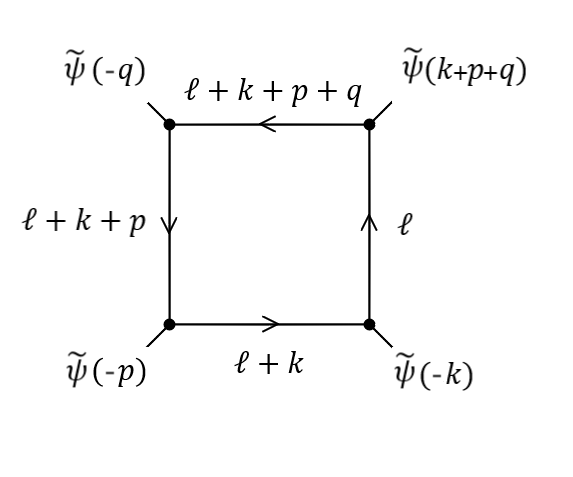}
     \caption{Scalar box integral for the contribution $\Gamma_C$.}
     \label{fig:Box}
\end{figure}

This integral, with different powers of the propagators, is known as the
scalar box integral, illustrated in Fig. \ref{fig:Box}. It has been dealt
with applying different mass conditions on the internal and external
momentum lines, as well as in the on-shell or off-shell conditions for
them, for example in \cite{hooft1979integrals}, \cite{duplancic2001box} and
\cite{haug2023box}. However, in our case the kernel is defined trough
negative and rational powers of the propagators, which gives divergent
integrals when using Feynman parametrization in \eqref{K definition}. 
For this reason, we procceed to solve the integral by imposing a specific form
for $\psi(x_\shortparallel)$.
\subsection{Wave-like surface}
We can obtain a result for \eqref{K definition} if the surface takes the form 
\begin{equation} \label{Static surface definition}
    \psi(x_\shortparallel)=2A\, {\rm cos}(\omega_0 \, x_0 - \bm{\omega}_\shortparallel \bm{x}_\shortparallel),
\end{equation}
defining the vector $\omega_\shortparallel=(\omega_0,\bm{\omega}_\shortparallel )$ where $\bm{\omega}_\shortparallel$ and $\bm{x}_\shortparallel$ have their components ranging from $1$ to $d-1$. Then we have
\begin{align} \label{Dynamic surface Fourier transform powers}
\tilde{\psi}(k_\shortparallel)&=A\,(2\pi)^d\left[\delta^d(k_\shortparallel-\omega_\shortparallel)+\delta^d(k_\shortparallel+\omega_\shortparallel)\right] \\ \nonumber
\tilde{\psi^2}(k_\shortparallel)&=A^2(2\pi)^d\left[\delta^d(k_\shortparallel-2\omega_\shortparallel)+\delta^d(k_\shortparallel+2\omega_\shortparallel)+2\delta^d(k_\shortparallel)\right] \\ \nonumber
\tilde{\psi^3}(k_\shortparallel)&= A^3(2\pi)^d\left[\delta^d(k_\shortparallel-3\omega_\shortparallel)+\delta^d(k_\shortparallel+3\omega_\shortparallel)+3\delta^d(k_\shortparallel-\omega_\shortparallel) + 3\delta^d(k_\shortparallel+\omega_\shortparallel)\right],
\end{align}
where we see the tendency that the $n$-th power of the expansion involves a $n$-multiple of the wave vector.
This leads to
\begin{align} \label{Particular result k=2 n=2 C}
    \Gamma_C&= \int_{\not{k_\shortparallel}\,\not{p_\shortparallel}\,\not{q_\shortparallel}} \tilde{\psi}(-k_\shortparallel) \tilde{\psi}(-p_\shortparallel) \tilde{\psi}(-q_\shortparallel) \tilde{\psi}(k_\shortparallel+p_\shortparallel+q_\shortparallel) \, K(k_\shortparallel,p_\shortparallel,q_\shortparallel) \\ \nonumber
    &=A^4(2\pi)^{2d}\\ \nonumber
    &\times\int_{\not{k_\shortparallel}\,\not{p_\shortparallel}\,\not{q_\shortparallel}}\left[\delta^d(k_\shortparallel-\omega_\shortparallel)+\delta^d(k_\shortparallel+\omega_\shortparallel)\right]  \left[\delta^d(p_\shortparallel-\omega_\shortparallel)+\delta^d(p_\shortparallel+\omega_\shortparallel)\right]  \\ \nonumber
    &\times \left[\delta^d(q_\shortparallel-\omega_\shortparallel) +\delta^d(q_\shortparallel+\omega_\shortparallel)\right] \left[\delta^d((k_\shortparallel+p_\shortparallel+q_\shortparallel)-\omega_\shortparallel)+\delta^d((k_\shortparallel+p_\shortparallel+q_\shortparallel)+\omega_\shortparallel)\right]\\ \nonumber &\times  \int_{\not{\ell}_\shortparallel} |\ell_\shortparallel| |\ell_\shortparallel+k_\shortparallel| |\ell_\shortparallel+k_\shortparallel+p_\shortparallel| |\ell_\shortparallel+k_\shortparallel+p_\shortparallel+q_\shortparallel| \\ \nonumber
    &=4A^4(2\pi)^d\delta^d(0) \int \limits_{\not{\ell_\shortparallel}} |\ell_\shortparallel|^2 |\ell_\shortparallel-\omega_\shortparallel| |\ell_\shortparallel+\omega_\shortparallel| + \text{real term,}
\end{align}
where the loop integral \eqref{K definition} is now $H(\omega_\shortparallel,-2\omega_\shortparallel)$, given by \eqref{H loop integration}. Similarly to \eqref{Development 4 k=2 n=2}, we discarded a term proportional to $|\ell_\shortparallel|^2 |\ell_\shortparallel+\omega_\shortparallel|^2$ since it doesn't contribute to the imaginary part. 

We note that the factor $(2 \pi)^d \delta^d(0)$ appears, as expected, for a quantity that has been calculated for an infinite region of space-time. This corresponds to the integration hypervolume $T\,V$. This factor will appear on all contributions and can be made sense of by dividing every term by it, getting as a result the respective propabilities per unit time and per spatial volume.

Now using \eqref{Dynamic surface Fourier transform powers} on the previous calculations, we find
\begin{align}
    \Gamma_\Delta^{(4,1)}&= \frac{\Gamma(-2-\frac{d}{2}) \, \Gamma(\frac{d+3}{2}) \Gamma(\frac{d+1}{2})}{(4\pi)^{\frac{d}{2}} \Gamma(-\frac{3}{2}) \Gamma(-\frac{1}{2}) \Gamma(d+2) } \int_{\not{k_\shortparallel}} \left( \frac{\tilde{\psi^2}(k_\shortparallel)\tilde{\psi^2}(-k_\shortparallel)}{8} - \frac{\tilde{\psi^3}(k_\shortparallel) \tilde{\psi}(-k_\shortparallel)}{6}\right) \left(k_\shortparallel^2\right)^{\frac{d}{2}+2} \\ \nonumber
    &=\frac{\Gamma(-2-\frac{d}{2}) \, \Gamma(\frac{d+3}{2}) \Gamma(\frac{d+1}{2})}{(4\pi)^{\frac{d}{2}} \Gamma(-\frac{3}{2}) \Gamma(-\frac{1}{2}) \Gamma(d+2) } A^4 \delta^d(0) (2 \pi)^d \left( \frac{1}{4}((2 \omega_\shortparallel)^2)^{\frac{d}{2}+2} - ((\omega_\shortparallel)^2)^{\frac{d}{2}+2} \right) \\ \nonumber
    &=\frac{A^4 \delta^d(0) (2 \pi)^d  (\omega_\shortparallel^2)^{{\frac{d}{2}+2}}}{(4\pi)^{\frac{d}{2}}} \, \Gamma(-2-\frac{d}{2})  \, (4^{\frac{d+2}{2}}-1) \, a 
\end{align}
and
\begin{align}
    \Gamma_\Delta^{(4,2)}&=-\frac{1}{16} \Gamma_A + \frac{1}{2} \Gamma_B - \frac{1}{4} \Gamma_C \\ \nonumber
    &= \frac{A^4 (2 \pi)^d \delta^d(0) (\omega_\shortparallel^2)^{\frac{d}{2}+2}}{(4 \pi)^{\frac{d}{2}}} \\ \nonumber
    &\times \left[ \Gamma(-2-\frac{d}{2}) \left(-4^{\frac{d+2}{2}}\,a +d ( \frac{3}{2}-4^{\frac{d+3}{2}}) \, b\right) + \Gamma(-1-\frac{d}{2}) \left( c - 4^{\frac{d+2}{2}} e\right) \right] 
\end{align}
with the definitions:
\begin{align}
    &a=  \frac{\Gamma(\frac{d+3}{2}) \Gamma(\frac{d+1}{2})  }{\Gamma(-\frac{3}{2}) \Gamma(-\frac{1}{2}) (d+1)!} \\ \nonumber
    &b= \frac{ \left( \Gamma(\frac{d+3}{2}) \right)^2 }{\left(\Gamma(-\frac{1}{2})\right)^2 (d+2)!} \\ \nonumber
    &c= \frac{3 \left(\Gamma(\frac{d+1}{2})\right)^2}{d\,! \left(\Gamma(-\frac{1}{2})\right)^2} + \frac{3 \Gamma(\frac{d+1}{2}) \Gamma(\frac{d+5}{2})}{\left(\Gamma(-\frac{1}{2})\right)^2(d+2)!} - \frac{2 \Gamma(\frac{d+1}{2}) \Gamma(\frac{d+3}{2})}{\left(\Gamma(-\frac{1}{2})\right)^2(d+1)!} \\ \nonumber
    &e= \frac{ \left(\Gamma(\frac{d+1}{2})\right)^2}{d\,! \left(\Gamma(-\frac{1}{2})\right)^2} + \frac{4 \Gamma(\frac{d+1}{2}) \Gamma(\frac{d+5}{2})}{\left(\Gamma(-\frac{1}{2})\right)^2(d+2)!} - \frac{4 \Gamma(\frac{d+1}{2}) \Gamma(\frac{d+3}{2})}{\left(\Gamma(-\frac{1}{2})\right)^2(d+1)!} \;.
\end{align}
All these constants can be obtained directly for even and odd dimensions by plugging the value of $d$ without any extra work, since they're all well defined. 
Therefore

\begin{align}
    \Gamma_\Delta^{(4)}&=\Gamma^{(4,1)}_\Delta + \Gamma^{(4,2)}_\Delta \\ \nonumber
    &= \frac{A^4 (2 \pi)^d \delta^d(0) (\omega_\shortparallel^2)^{\frac{d}{2}+2}}{(4 \pi)^{\frac{d}{2}}} \,  \\ \nonumber
    &\times \left[ \Gamma(-2-\frac{d}{2})  \left(-\,a +d ( \frac{3}{2}-4^{\frac{d+3}{2}}) \, b\right) + \Gamma(-1-\frac{d}{2}) \left( c - 4^{\frac{d+2}{2}} e\right) \right] .
\end{align}

We see that $\Gamma^{(4)}_\Delta$ has a common factor of $(\omega_\shortparallel^2)^{\frac{d}{2}+2}$ where, following the logic we used in sections \ref{Order 2 odd d} and \ref{Order 2 even d}, we identify that for odd $d$ there's a contribution to the imaginary part that comes from $\sqrt{\bm{\omega}_\shortparallel^2-\omega_0^2} \, \theta(|\omega_0| - |\bm{\omega}_\shortparallel|)$, and for even $d$, using dimensional regularization with $d=(2q-\epsilon)$, there's a contribution to the imaginary part coming from $\log({\bm{\omega}_\shortparallel^2-\omega_0^2}) \, \theta(|\omega_0| - |\bm{\omega}_\shortparallel|)$.

\begin{figure}[h] 
     \centering
     \includegraphics[width=0.7\textwidth]{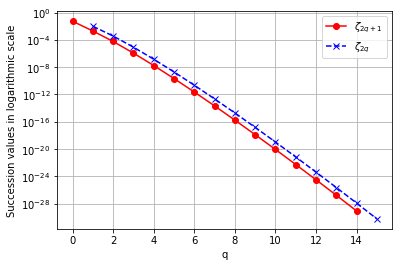}
     \caption{Successions $\zeta_{2q+1}$ and $\zeta_{2q}$ in logarithmic scale.}
     \label{fig:ratio_order4}
\end{figure}

Writing the imaginary part of $\Gamma^{(4)}$, the real time effective action corresponding to $\Gamma^{(4)}_\Delta$, as

\begin{equation}
    {\rm Im}(\Gamma^{(4)})= A^4 (2 \pi)^d \delta^d(0) \, \zeta_{2q+1} \, \left( (\omega^0)^2-\bm{\omega}_\shortparallel^2 \right)^{q+2+\frac{1}{2}}\theta(|\omega^0| - |\bm{\omega}_\shortparallel|) \, \,\, 
\end{equation}
for $d=2q+1$ and
\begin{equation}
    {\rm Im}(\Gamma^{(4)})= A^4 (2 \pi)^d \delta^d(0) \, \zeta_{2q} \, \left( (\omega^0)^2-\bm{\omega}_\shortparallel^2 \right)^{q+2} \theta(|\omega^0| - |\bm{\omega}_\shortparallel|)
\end{equation}
for $d=2q$, we get




\begin{align}
&\zeta_{2q+1}=\frac{(-1)^{q}}{(4\pi)^{q+\frac{1}{2}}} \times \\ \nonumber
&\left[ \Gamma(-q-\frac{3}{2}) (c-4^{q+\frac{3}{2}}e) + \Gamma(-q-\frac{5}{2}) \left( - \,a + (2q+1) (\frac{3}{2}-4^{q+2}) b \right) \right],
\end{align}
and

\begin{align}
   \zeta_{2q}&= \frac{\pi}{(4\pi)^{q}} \left[\frac{(c-4^{q+1}e)}{(q+1)!}   -\frac{\left( -\, a + 2q (\frac{3}{2}-4^{q+\frac{3}{2}}) b \right)}{(q+2)!}  \right].
\end{align}
The successions $\zeta_{2q+1}$ and $ \zeta_{2q}$ are plotted in Fig. \ref{fig:ratio_order4}. Just like for the second order contribution, these successions are positive and decrease rapidly with growing dimensions, showing a linear behaviour on logarithmic scale.

\section{Conclusions}\label{sec:conc}
We have evaluated the imaginary part of the effective action, and therefore the corresponding probability of vacuum decay, for a Dirichlet surface in $d+1$ dimensions that can deform and move, in a time-dependent way, under the assumtion of small departures with respect to an average, planar hypersurface. 
This evaluation has been performed up to the fourth order in the amplitude of the deformation, giving rather explicit expressions for the second order term; while the fourth order one is presented for the case of wave-like deformations of the
surface. 

In all the terms we have evaluated, there is a common threshold for vacuum decay, namely, that the deformation should have time-like components in Fourier space.
Note that, given the (assumed) bounded nature of the deformations, this kind of motion should corresspond, at least locally at each point of the surface, to some sort of oscillatory motion.

We have presented rather general results regarding their dependence on the number of spatial dimensions $d$. The general structure of the decay probability per unit time and per unit volume, does have a dependence on the momenta of the deformation, which may be explained on dimensional grounds.  On the other hand, these general results have a dimensionless prefactor, of which we know the dependence with $d$, for which we find a consistent exponential decay when the number of dimensions increases. This means that the process of pair creation by a
surface of codimension one seems to be less effective (at least per unit volume)
when $d$ becomes larger. The total probability, of course, does have an exponentially growing factor $L^d$.

\end{document}